\documentstyle[prb,aps,epsf,array,amssymb]{revtex}

\def\xor{\oplus}
\input{psfig.sty}

\begin{document}
\twocolumn[\hsize\textwidth\columnwidth\hsize\csname @twocolumnfalse\endcsname

\title{\bf Long-time dynamics of de Gennes' model for reptation}

\author{G. T. Barkema}

\address{HLRZ, Forschungszentrum J\"ulich GmbH, D-52425 J\"ulich, Germany,
and \\
ITP, Utrecht University, Princetonplein 5, 3584 CC Utrecht, the Netherlands
\cite{barkadd}}

\author{H. M. Krenzlin}

\address{Institut f\"ur Festk\"orperforschung,
Forschungszentrum J\"ulich GmbH, D-52425 J\"ulich, Germany}

\maketitle

\begin{abstract}
Diffusion of a polymer in a gel is studied within the framework of de
Gennes' model for reptation.  Our results for the scaling of the
diffusion coefficient $D$ and the longest relaxation time $\tau$ are
markedly different from the most recently reported results, and are in
agreement with de Gennes' reptation arguments: $D \sim N^{-2}$ and
$\tau \sim N^3$.  The leading exponent of the finite-size corrections
to the diffusion coefficient is consistent with the value of -2/3 that
was reported for the Rubinstein model.  This agreement suggests that
its origin might be physical rather than an artifact of these models.
\end{abstract}
]
\narrowtext

In the theory of reptation as formulated by de Gennes \cite{deG} it is
predicted that the asymptotic behavior of the diffusion coefficient $D$
of an entangled polymer scales with the chain length as $D \sim N^{-2}$,
and that the longest relaxation time $\tau$ scales as $\tau \sim N^3$.
Evans and Edwards \cite{EE} were the first to present numerical studies
of de Gennes' reptation model, which describes a single reptating polymer,
and their results were in agreement with
theoretical expectations. Recent simulations of the short-time
behavior of this model are also in agreement with reptation
arguments\cite{baumg}.
However, the simulations of the long-time behavior of this model by
Deutsch and Madden \cite{dm} contradict de Gennes' predictions with,
instead, $D\sim N^{-2.5 \pm 0.04}$ and $\tau \sim N^{3.41 \pm 0.14}$.

This paper presents simulation results on the same model as {{that}}
studied by Deutsch and Madden. The advance in computers combined with
a highly optimized multispin coding technique allows us to simulate
over time scales that are several orders of magnitude longer than
previous simulations, yielding more accurate data points for longer chains,
which in turn allows for a data analysis that includes corrections to the
leading power-law behavior.

An alternative model for reptating polymers was proposed by Rubinstein
\cite{Rubinstein}, and has been studied numerically by one of us
\cite{repton,Drepton}.  It was found that the diffusion coefficient $D$
scales as $D \sim N^{-2}$, and the exponent of the leading finite-size
correction was found to be close to $-2/3$.  In an analytical work by
Pr\"ahofer and Spohn \cite{prae} the $N^{-2}$ behavior was derived
rigorously with a variational method, and the leading exponent of
finite-size corrections was shown to be between $-1/2$ and $-1$, with the
former the more likely. This paper will show that de Gennes' model
for reptation and Rubinsteins repton model have the same
length-dependence of the diffusion coefficient, both the leading and the
subleading part in the long-length limit; such agreement between two
different models enhances the credibility that these results are physical.

In this paper, we will first describe the model considered and then
motivate the computational method used for the simulations.  This part
will be followed by the calculation of the diffusion constant and the
results for the longest relaxation times.

\section{de Gennes' model for reptation}

We simulate a model for a polymer in a gel that was introduced by de
Gennes.  In this model, a polymer of length $N$ is represented by
$N+1$ monomers on {{a 3D cubic}} lattice, connected by a sequence of
$N$ steps on lattice edges. One elementary move in this model is made
by randomly selecting one monomer, and attempting to move it. If this
monomer is an endpoint of the polymer, or if its two steps are located
on the same edge, it will randomly move to one of the six possible
lattice sites (the site it is already in, and five new ones).  The
time scale involved with one such an elementary move is
$\Delta t=1/(6N+6)$, so that the rate of any monomer to move to any
allowed lattice site is one time unit.

A configuration of the polymer in the two-dimensional version of this
model is illustrated in Figure 1. In this configuration,
monomers 1, 5, 11, 12, and 15 can move to three other locations, the
other monomers are frozen.  Note that all our simulations were done in
the three-dimensional model.

\begin{figure}
\epsfxsize=8cm
\epsfbox{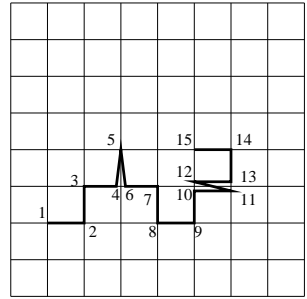}
\caption{Two-dimensional version of de Gennes' model for a reptating
polymer with length $N=14$ steps.}
\label{eemodel}
\end{figure}

\section{simulation method}

Rather than storing the spatial coordinates all along the chain, it is
for simulation purposes more convenient to store the spatial
coordinates of only the endpoints, and the sequence of steps from one
of them to the other.

In one elementary move in a direct implementation of the dynamics, the
first step, the last step, or a pair of neighbouring steps is randomly
selected and proposed to move.  For the first and last step, this
involves simply a replacement by a random step; a pair of steps in
the middle of the chain can only move if the steps are opposite, and in
that case it is replaced by a randomly selected pair of opposite steps.
Every possible change in the configuration should be proposed once per
time unit, and therefore $6N+6$ such elementary moves should be
performed per unit of time. As we will see in section \ref{tauresults},
the typical correlation time is approximately $0.13N^3$, and thus around $N^4$
elementary moves have to be performed to obtain a statistically
independent configuration. Finite-size effects disappear only if $N$
is well above 100, requiring arond $10^8$ elementary moves per
statistically independent configuration; an efficient implementation
is crucial for numerical accuracy.

To achieve such a high efficiency, we use {\it multispin coding}.
Most of our results were obtained on a DEC alpha workstation with a
64-bit processor, allowing 64 simulations to be run in
parallel. For each of the 64 simulations, we stored the coordinates of
both endpoints as $x_l^{(0)} \dots x_l^{(63)}$ for the left ends, and
$x_r^{(0)} \dots x_r^{(63)}$ for the right ends. The steps $0 \dots N-1$
are stored in $3N$ long integers of 64 bits. Suppose that we
denote the $k^{th}$ bit of long integer $w$ as $w^k$, then step $i$ of
simulation $k$ is stored in the three bits $\{a[i]^k, b[i]^k, c[i]^k\}$;
if the step is in the positive $x$-, $y$-, or
$z$-direction, this triplet is equal to \{1,0,0\}, \{0,1,0\} or
\{0,0,1\}, respectively; the negative directions are represented by
\{0,1,1\}, \{1,0,1\}, and \{1,1,0\}, respectively. Note that opposite
steps are each other's binary complement.

To determine whether a hairpin is located in configuration $k$ at steps
$i$ and $j=i+1$, we have to check whether the triplets
$\{a[i]^k, b[i]^k, c[i]^k\}$ and $\{a[j]^k, b[j]^k, c[j]^k\}$
are complementary. This can be done with the following logical statement:

\begin{equation}
h=(a[i] \xor a[j]) \land (b[i] \xor b[j]) \land (c[i] \xor c[j]);
\end{equation}

the $k^{th}$ bit of long integer $h$ is one if there is a hairpin,
otherwise it is zero. With this {\it one} statement, we have however
determined the location of a hairpin in {\it all} 64 simulations! This
is the power of multispin coding.

Of course it is not sufficient to determine the presence of a hairpin,
we also have to replace hairpins by a random pair of opposite steps.
Suppose we have available a triplet of long integers \{ra, rb, rc\}
that represent 64 random steps, then the logical statement

\begin{equation}
a[i]=(h \land ra) \lor (\lnot h \land a[i]);
\end{equation}

and equivalent statements for b[i], c[i], a[j], b[j], and c[j] suffice
to make one elementary move in all 64 simulations. In addition to
these 24 logical operations some load- and store operations are
required; altogether, we can make 64 elementary moves (one per
simulation) in a fraction of a microsecond.

The endpoints require more care, since they involve keeping track of
their spatial coordinates. For these, we needed a loop over the 64
simulations.

Since we perform long simulations, the generation of random numbers is
a major concern. In our simulation, there are two places where random
numbers are used: first, in the selection of the step $i$, secondly in
generating a random triplet of long integers \{ra, rb, rc\} that
represent 64 random steps.  For the first, we used the additive lagged
Fibonacci generator proposed by Mitchell and Moore \cite{mm}. For the
second, we initially generated a set of $M$ such triplets by building
them up bit- by-bit with the Mitchell-Moore generator, and after using
this set, a new set was generated from it by three operations:
\begin{itemize}
\item{} each triplet was permutated with 50\% probability
\item{} each triplet was inverted with 50 \% probability
\item{} the list was reshuffled randomly.
\end{itemize}
For the first and second operation, we used a random bit generator
proposed by Marsaglia \cite{marsag}. Note that we do not rely on the spectral
properties of this generator, nor on the quality of each individual bit
in the Mitchell-Moore generator. (Nor do we claim imperfections in these
generators.)

\section{The diffusion constant}

To determine the diffusion constant $D$, we started with 64 random
polymer configurations, and let them evolve over a long time using the
multispin approach outlined above. The coordinates $\vec{r}_l$ and
$\vec{r}_r$ of the endpoints of each polymer were stored at regular
time intervals. From their average $\vec{r}_a \equiv (\vec{r}_l+\vec{r}_r)/2$
we calculate the function
\begin{equation}
D(t) = \frac{\langle (\vec{r}_a(t_0+t)-\vec{r}_a(t))^2 \rangle -N/4}{6t};
\label{Deq}
\end{equation}
the long-time limit of this function gives the diffusion coefficient:
\begin{equation}
D=\lim_{t \rightarrow \infty} D(t).
\end{equation}
The additional term $N/4$ in Eq. (\ref{Deq}) corrects for the rapid
fluctuations of the endpoints relative to the center of mass, that do
not contribute in the long-time limit; it reduces the effects of the
finiteness of the time interval significantly. As we will see in
section \ref{tauresults}, the typical relaxation time scales as $\tau
\sim N^3$; we find that the long-time limit of $D(t)$ is well
reached at $t=6N^3$, and used this time interval to measure $D$.
Figure 2 illustrates the convergence of $D(t)$ as a function of
$t$.

\begin{figure}
\epsfxsize=8cm
\epsfbox{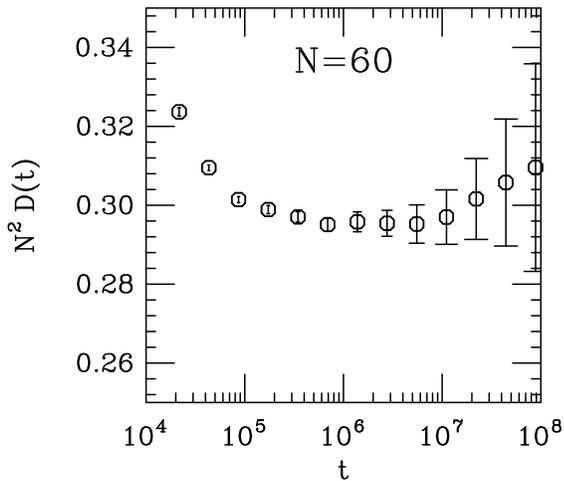}
\caption{Finite-time estimate $N^2 D(t)$ for the rescaled diffusion
constant $DN^2$ as a function of the length $t$ of the time interval,
for a polymer of length $N=60$.
\label{D60}}
\end{figure}

The results for the diffusion coefficient $D$ as a function of
length $N$ are given in Table I. The length of the runs
depends on the chain length $N$, and is also given in Table I.
For $N=150$ and $N=200$, we used a 32-bit computer, all
other results were obtained on a 64-bit computer.

\begin{figure}
\epsfxsize=8cm
\epsfbox{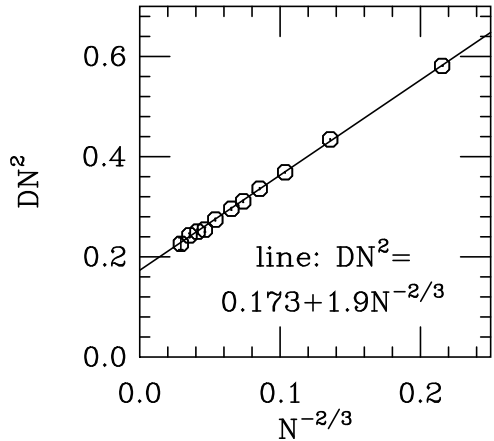}
\caption{Rescaled diffusion coefficient $DN^2$ as a function of $N^{-2/3}$.
The straight line is given by equation \ref{Dfit}.
\label{D}}
\end{figure}

In figure \ref{D}, the measured values for $DN^2$ are plotted as a function
of $N^{-2/3}$; all measurements fall on a single straight line given by
\begin{equation}
DN^2 = 0.173 + 1.9 N^{-2/3}.
\label{Dfit}
\end{equation}

\begin{table}[htb]
\caption{The diffusion constant $D$ times the squared length, for
polymers of several lengths $N$. The third and fourth columns give the
number of simulations and the length of the runs.}
\begin{center}
\begin{tabular}{cccc}
$~~~~N~~~~$ & $~~DN^2~~$  & \#runs & $t$ \\
\hline
\multicolumn{2} {c} {}\\
10    & 0.581 (0.001)  & 64 & $3.2 \cdot 10^5$\\
20    & 0.434 (0.001)  & 64 & $2.6 \cdot 10^6$\\
30    & 0.369 (0.001)  & 64 & $8.6 \cdot 10^6$\\
40    & 0.336 (0.002)  & 64 & $4.1 \cdot 10^7$\\
50    & 0.311 (0.002)  & 64 & $8.0 \cdot 10^7$\\
60    & 0.296 (0.003)  & 64 & $1.4 \cdot 10^8$\\
80    & 0.275 (0.002)  & 64 & $3.3 \cdot 10^8$\\
100   & 0.254 (0.005)  & 64 & $3.2 \cdot 10^8$\\
120   & 0.251 (0.005)  & 64 & $5.5 \cdot 10^8$\\
150   & 0.243 (0.004)  & 32 & $6.8 \cdot 10^8$\\
200   & 0.226 (0.010)  & 32 & $2.8 \cdot 10^8$\\
\end{tabular}
\end{center}
\label{Dtable}
\end{table}

This indicates that for long polymers, the asymptotic behavior of the
diffusion is in agreement with reptation predictions of de Gennes:  $D
\sim N^{-2}$. Also, it gives strong indication that the leading exponent
of the finite-size corrections has a non-integer value close to -2/3.
This exponent has also been observed in the Rubinstein repton model
\cite{repton,Drepton}, another lattice model for diffusing polymers,
and makes it plausible that it is not simply an artifact of the model.

\section{the longest relaxation times}
\label{tauresults}

\begin{figure}
\epsfxsize=8cm
\epsfbox{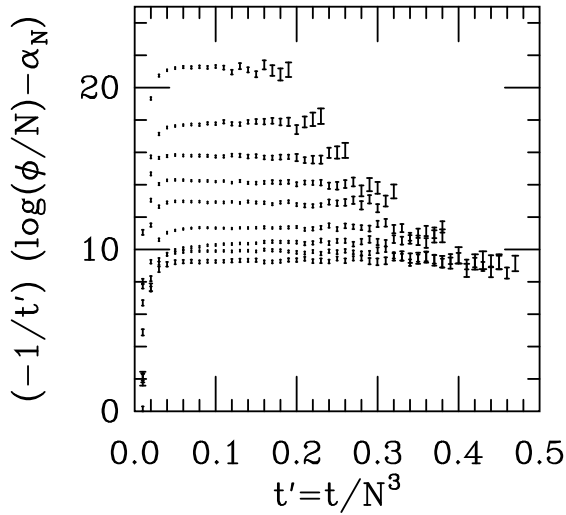}
\caption{ For several values of $N$, we have plotted
$(-1/t')(\log(\phi(t)/N-\alpha_N)$ as a function of $t'=t/N^3$.  All
curves show convergence to a constant value, which is equal to
$N^3/\tau_N$. From top to bottom, the curves
correspond to $N=20,$ 30, 40, 50, 60, 80, 100, 120, and 150.
\label{curall}}
\end{figure}

The shape of the polymer at time $t_0+t$ stays correlated to its shape
at time $t_0$ for a long time, but eventually decays exponentially
fast.  This final decay sets the longest relaxation time; it is often
related to the viscosity.  The scaling of the longest relaxation time
as a function of length is the topic of this section.  To determine the
longest relaxation time, we study the decay in the orientation
$\vec{d}=\vec{r}_r-\vec{r}_l$. The function $\phi(t)$ is the two-time
autocorrelation of this vector:
\begin{equation}
\phi(t)=\langle \vec{d}(t_0) \cdot \vec{d}(t_0+t) \rangle.
\end{equation}
We know that $\phi(0)=N$, since the expected end-to-end distance of the
polymer is $\sqrt{N}$. The function $\phi(t)$ is expected to decay
exponentially to zero as a function of $t$, with a typical length-dependent
relaxation time $\tau_N$:
\begin{equation}
\phi(t) \approx N \exp(-t/\tau_N).
\end{equation}
To determine the correlation time $\tau_N$, we have plotted in Figure
\ref{curall} $(-1/t')(\log(\phi(t)/N-\alpha_N)$ as a function of
$t'=t/N^3$; ${\alpha}_N$ is an $N$-dependent parameter that removes
partly the finite-time corrections but does not affect the scaling at
large $t'$:  since $\log(\phi(t))$ grows roughly linearly with $t$, it
is the dominant term for longer $t$. The values for ${\alpha}_N$ range
from ${\alpha}_{20} = -0.62$ to ${\alpha}_{150} = -0.44$.  The
asymptotic values of the curves in figure \ref{curall} are equal to
$N^3/\tau_N$.

The values for $N^3/\tau_N$ that we obtained in this way are given in
Table II, as well as their reciprocals; the errors are our estimates by
eye.

\begin{table}[htb]
\caption{The longest relaxation time $\tau_N$ divided by $N^3$,
for polymers of several lengths $N$.}
\begin{center}
\begin{tabular}{c@{\hspace{2cm}}c@{\hspace{2cm}}c}
$N$ & $N^3/\tau_N$ & $\tau_N/N^3$ \\
\hline
20    & 21.2 (0.2) & 0.0472 (5)\\
30    & 18.0 (0.2) & 0.0556 (6)\\
40    & 15.7 (0.2) & 0.0637 (8)\\
50    & 14.2 (0.2) & 0.070 (1)\\
60    & 12.9 (0.2) & 0.078 (1)\\
80    & 11.3 (0.2) & 0.088 (2)\\
100   & 10.4 (0.2) & 0.096 (2)\\
120   & 9.8 (0.2) & 0.102 (2)\\
150   & 9.3 (0.2) & 0.107 (2)\\
\end{tabular}
\end{center}
\label{tautable}
\end{table}

\begin{figure}
\epsfxsize=8cm
\epsfbox{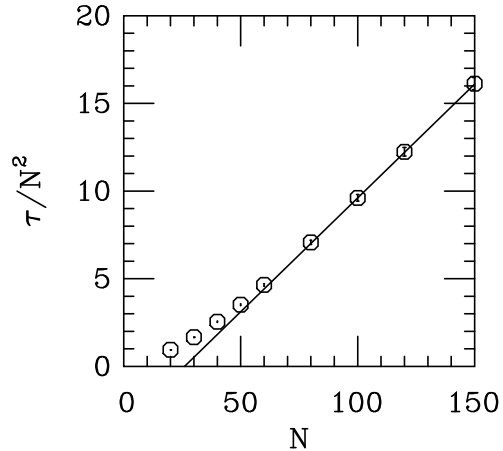}
\caption{The longest relaxation time $\tau_N$ divided by $N^2$, as a function
of $N$. The straight line is a fit given by $\tau_N=0.13N^2+c$, indicating that
asymptotically $\tau \approx 0.13N^2$.
\label{tau}}
\end{figure}

It is expected from reptation arguments that in the limit of infinitely
long chains, the product $D\tau$ scales linearly with $N$; as we found in
the previous section that for long chains $D \sim N^{-2}$, this would mean
that for long chains $\tau \sim N^3$. To check this, we have
plotted $\tau/N^2$ as a function of $N$ in figure \ref{tau}. For large $N$, the
measurements are fitted well by a straight line with slope 0.13,
indicating that for long polymers $\tau \approx 0.13 N^3$.

\section{Conclusions}
We have performed numerical simulations of de Gennes' model for
polymer reptation. For the diffusion coefficient $D$, we find
$D=N^{-2} (0.173+1.9N^{-2/3})$. For the longest relaxation
time $\tau$ of long polymers we find $\tau\approx 0.13 N^3$. The
leading-orders of the diffusion coefficient and the longest
relaxation time are in agreement with theoretical arguments of de
Gennes.  There is no theoretical prediction for the leading exponent of
the finite-size corrections for the diffusion coefficient; our value of
-2/3 is identical to earlier reports of this exponent in the Rubinstein
repton model.

\section{Acknowledgements}

We acknowlege useful and interesting discussions with Artur Baumg\"artner,
Klaus Kehr and Mark Newman.

\bibliographystyle{prsty}

\end{document}